\documentclass[aps, prb, twocolumn, showpacs, superscriptaddress]{revtex4-1}

\usepackage{amsmath,amssymb}
\usepackage{graphicx}
\usepackage{dcolumn}
\usepackage{bm}
\begin{document}

\title {Absence of coherent peaks in a $Z_{2}$ fractionalized BCS superconducting state}
\author{Yin Zhong}
\email{zhongy05@hotmail.com}
\affiliation{Center for Interdisciplinary Studies $\&$ Key Laboratory for
Magnetism and Magnetic Materials of the MoE, Lanzhou University, Lanzhou 730000, China}
\author{Han-Tao Lu}
\affiliation{Center for Interdisciplinary Studies $\&$ Key Laboratory for
Magnetism and Magnetic Materials of the MoE, Lanzhou University, Lanzhou 730000, China}
\author{Hong-Gang Luo}
\email{luohg@lzu.edu.cn}
\affiliation{Center for Interdisciplinary Studies $\&$ Key Laboratory for
Magnetism and Magnetic Materials of the MoE, Lanzhou University, Lanzhou 730000, China}
\affiliation{Beijing Computational Science Research Center, Beijing 100084, China}

\date{\today}

\begin{abstract}
We explore a $Z_{2}$ fractionalized Bardeen-Cooper-Schrieffer (BCS) superconducting state, which is a minimal extension of usual BCS framework. It is found that this state has similar thermal and transport properties, but its single-particle feature strongly deviates from the coherent quasiparticle behavior of the classic/conventional BCS superconducting state. The fingerprint of such $Z_{2}$ BCS state is the absence of the BCS coherent peaks and instead a kink in the local density of state occurs, which in principle could be probed by scanning tunneling microscopy or point-contact spectroscopy experiments. The corresponding exactly soluble models that realize the desirable $Z_{2}$ fractionalized BCS state is presented. In addition, we also study the extended $t$-$U$-$J$ model by using $Z_{2}$ slave-spin representation and find that the $Z_{2}$ BCS state may exist when the paring structure is fully gapped or has nodes. The prototypical wave-function of such a $Z_{2}$ BCS state is also proposed, which could be taken as trial wave-function in current numerical techniques. Furthermore, the pairing mechanism of $Z_{2}$ BCS state is argued from both weak and strong coupling perspective. The present work may be helpful to further study the unconventional superconductivity and its relation to non-Fermi liquids.
\end{abstract}

\maketitle

\section{Introduction} \label{intr}
Unconventional superconductivity\cite{Sigrist1991,Scalapino2012} is now ubiquitous in strongly correlated electron systems, such as
cuprate\cite{Lee2006,Armitage2010}, pnictide\cite{Stewart2011,Chubukov2012} and heavy fermion compound\cite{Steglich2005}. However, theoretical understanding of those unconventional superconductors and its underlying pairing mechanism are still rather uncompleted despite extensive and intensive study in the past decades. The common belief of superconductivity is that the superconducting state results from the pairing instability of original non-superconducting normal state. This is implicitly assumed in the classic Bardeen-Cooper-Schrieffer (BCS) theory where the usual Fermi liquid state plays the role of the non-superconducting background state.\cite{Sachdev2011} Thus, logically and in principle a non-Fermi liquid normal phase might lead to a corresponding unconventional superconducting phase if exists. Recently, the simplest non-Fermi liquid called orthogonal metal, which has similar thermal and transport properties of the conventional Landau Fermi liquid but with gapped single-particle excitation, has proposed based on the study of $Z_{2}$ slave-spin representation.\cite{Nandkishore,Ruegg2012} Some extensions have been investigated in detail by mean-field theory or by exactly soluble models constructed.\cite{Zhong2012e,Zhong2012,Zhong2012,Zhong2013b,Zhong2013c,Ruegg2013,Maciejko2013}

In light of those minimal non-Fermi liquid metallic states, we explore a $Z_{2}$ fractionalized BCS superconducting state, which could be considered as a minimal extension of usual BCS framework. It is found that this state has similar thermal and transport properties but its single-particle feature strongly deviates from the coherent quasiparticle behavior in the usual BCS state. The signature of such $Z_{2}$ BCS state is the absence of the BCS coherent peak and instead a kink in the local density of state (DOS) occurs at the position of the coherent peaks. This result in principle could be probed by scanning tunneling microscopy\cite{Fischer2007,Hoffman2011} or point-contact spectroscopy experiments. We emphasize that this new feature has not been reported in existing literature and is the main finding of the present paper.

To make further insight into such a $Z_{2}$ BCS state, we present an exactly soluble model that realize the desirable $Z_{2}$ fractionalized BCS superconducting state. We find the superfluid density is consistent with the usual BCS state but the coherent peaks are absent. To contact with realistic lattice models in condensed matter physics, the $Z_{2}$ slave-spin representation\cite{deMedici,Ruegg} is applied to the extended Hubbard or $t-U-J$ model. \cite{Xiang2009,Laughlin2002,Laughlin2013} The proposed $Z_{2}$ BCS phase may be stable fixed points in the phase diagram if the pairing structure has full gap or a small number nodes at underlying Fermi surface. The usual uniform s-wave and $d_{x^{2}-y^{2}}$-wave fulfill such requirement and even chiral $p+ip$/$d+id$ are possible as well. In addition, the prototypical wave-function of such fractionalized BCS state is also proposed, which could be utilized by numerical techniques. The relations among $Z_{2}$ BCS states, $SC^{\ast}$\cite{Chowdhury2013} and parent orthogonal metals are also analyzed. Interestingly, we argue the weak and strong coupling pairing mechanism of $Z_{2}$ BCS state, which implies a link among Fermi liquid, orthogonal metal and gauge-theory description. Since the parent orthogonal metals and the proposed $Z_{2}$ BCS phase have not been found experimentally, we have to expect they may be realized in the exactly soluble models in terms of the versatile cold-atom simulations.\cite{Bloch} If such experiments are available, the theoretical predictions are the mentioned absence of coherent peaks, the occurrence of a kink in DOS and the normal behavior of superfluid density response.

The remainder of this paper is organized as follows.
In Sec.\ref{sec2}, we introduces an exactly soluble model to illustrate the idea of the $Z_{2}$ fractionalized BCS state and the local DOS and superfluid density are calculated. In Sec.\ref{sec3}, we study the extended Hubbard or $t-U-J$ model with the help of the $Z_{2}$ slave-spin representation. The interesting prototypical wave-function of the fractionalized BCS is given in Sec.\ref{sec4}.
Finally, we end this work with a brief conclusion in Sec.\ref{sec5}.

\section{The idea of $Z_{2}$ fractionalized BCS state and an exactly soluble model}\label{sec2}

\subsection{$Z_{2}$ fractionalized BCS state and the soluble model}
Firstly, let us consider the following exactly soluble model to illustrate the idea of the $Z_{2}$ fractionalized BCS state.[Extension to other lattices like honeycomb and triangular lattices is straightforward and we do not present it in this paper.]

\begin{eqnarray}
&&H=H_{K}+H_{c},\nonumber\\
&&H_{K}=-h\sum_{\langle ij\rangle}\hat{\sigma}^{z}_{ij}-J\sum_{i}\prod_{j=i+\delta_{a}}\hat{\sigma}^{x}_{ij}-W\sum_{i}\prod_{j\in \text{plaquett}}\sigma^{z}_{ij},\nonumber\\
&&H_{c}=\sum_{\langle ij\rangle\sigma}t_{ij}c_{i\sigma}^{\dag}\hat{\sigma}^{z}_{ij}c_{j\sigma}+\sum_{\langle ij\rangle}[\Delta_{ij}c_{i\uparrow}^{\dag}\hat{\sigma}^{z}_{ij}c_{j\downarrow}^{\dag}+\text{H.c.}],\label{eq1}
\end{eqnarray}
where $H_{K}$ describes an extended Kitaev toric code model\cite{Kitaev} on the square lattice and is dual to an unconstrained quantum transverse Ising model $H_{I}$ (see below). In $H_{K}$, $Z_{2}$ field $\hat{\sigma}^{z}_{ij},\hat{\sigma}^{x}_{ij}$ are usual Pauli matrices defined on the link of lattice and $h$-term serves as the kinetic energy while $J$-term acts like a potential ($j=i+\delta_{a}$ denotes four nearest-neighbor sites, i.e. $j=i\pm \hat{x},\pm \hat{y}$ ). $H_{c}$ presents the usual BCS mean-field Hamiltonian of conduction electrons $c_{i\sigma}$ coupled to the $Z_{2}$ field with $\Delta_{ij}$ being the pairing strength on lattice sites.

Following Refs.[\onlinecite{Nandkishore}] and [\onlinecite{Zhong2013c}], Eq.(\ref{eq1}) can be mapped to its dual formalism as
\begin{eqnarray}
H&&=H_{I}+H_{f},\nonumber\\
H_{I}&&=-h\sum_{\langle ij\rangle}\hat{\tau}^{x}_{i}\hat{\tau}^{x}_{j}-J\sum_{i}\hat{\tau}^{z}_{i},\nonumber\\
H_{f}&&=\sum_{\langle ij\rangle\sigma}t_{ij}f_{i\sigma}^{\dag}f_{j\sigma}+\sum_{\langle ij\rangle}[\Delta_{ij}f_{i\uparrow}^{\dag}f_{j\downarrow}^{\dag}+\text{H.c.}]\nonumber\\
&&=\sum_{k\sigma}\varepsilon_{k}f_{k\sigma}^{\dag}f_{k\sigma}+\sum_{k}[\Delta_{k}f_{k\uparrow}^{\dag}f_{-k\downarrow}^{\dag}+\text{H.c.}].\label{eq2}
\end{eqnarray}

Here, we have used the dual transformation $\prod_{j=i+\delta_{a}}\hat{\sigma}^{x}_{ij}=\hat{\tau}^{z}_{i}, \hat{\sigma}^{z}_{ij}=\hat{\tau}^{x}_{i}\hat{\tau}^{x}_{j}$ and
defined $c_{i\sigma}=\hat{\tau}_{i}^{x}f_{i\sigma}$. Thus, the $H_{f}$ is soluble and corresponds to the usual BCS Hamiltonian with pairing function $\Delta_{k}$. However, $f_{i\sigma}$ is not the original electron operator but just
the auxiliary fermion, so properties of physical electrons are determined by combined auxiliary fermion $f_{\sigma}$ (omitting the subscript) with the Ising spin $\hat{\tau}^{x}$.
For the quantum Ising model $H_{I}$, there exist two phases characterized by the order parameter $\langle \hat{\tau}^{x}_{i} \rangle=0$ when $h\ll J$ (the paramagnetic phase) and $\langle\hat{\tau}^{x}_{i}\rangle\neq0$ if $h\gg J$ (the ferromagnetic phase). Meanwhile, a second-order quantum phase transition exists between these two phases whose critical behaviors belong to the 3D Ising universal class.\cite{Sachdev2011}
Therefore, the ferromagnetic phase of Ising spin implies a usual BCS superconducting state since $c_{\sigma}\simeq\langle\hat{\tau}^{x}_{i}\rangle f_{\sigma}$ with nonzero quasiparticle weight $Z=\langle\hat{\tau}^{x}_{i}\rangle^{2}\neq0$. In contrast, the paramagnetic phase leads to a distinct BCS state because of coherence of electron and auxiliary fermion is lost due to gapped Ising spin $\langle \hat{\tau}^{x}_{i} \rangle=0$. We state that the system with $\langle \hat{\tau}^{x}_{i}\rangle=0$ still corresponds a BCS state since the auxiliary fermion $f_{\sigma}$ is always in the BCS state and it contributes physical thermal and transport
behaviors as usual BCS states.

In fact, the system with gapped Ising spins should be considered as a $Z_{2}$ fractionalized BCS state. This can be seen as follows.
Firstly, the extended Kitaev toric code model $H_{K}$ in Eq. (\ref{eq1}) is known to have the $Z_{2}$ gauge structure since the $Z_{2}$ gauge transformation operator $\hat{G}_{i}=\prod_{j\in \text{plaquett}}\sigma^{z}_{ij}$ is commuted with the Hamiltonian Eq.(\ref{eq1}).\cite{Zhong2013c} Then, when $h\ll J$, the dual Ising spin has $\langle \hat{\tau}^{x}_{i} \rangle=0$ and this relates to the vanishing small contribution of $h$ term in $H_{K}$. In other words, the ground-state should have $\prod_{j=i+\delta_{a}}\hat{\sigma}^{x}_{ij}=1$ and this is just the well-known deconfined state for the $Z_{2}$ gauge theory.\cite{Kogut} Therefore, the gapped Ising spin state is dual to the deconfined state of the $Z_{2}$ gauge theory.  Combined with the performed BCS state of $f$ fermions, we have obtained a $Z_{2}$ fractionalized BCS phase. Since the observable thermal and transport properties of such $Z_{2}$ BCS phase are dominated by the auxiliary fermion part and the Ising spin is gapped and incoherent in this phase, we expect the single particle feature, e.g., local density of state, should show strongly deviation from the coherent quasiparticle behavior in the usual BCS state. [Although the $Z_{2}$ fractionalized BCS state has the elusive topological order\cite{Wen2004,Senthil2000b} in contrast to the usual BCS one, this property is rather hard to be detected by current experimental techniques.]

\subsection{The local DOS of physical electrons for $Z_{2}$ fractionalized BCS state}

The local DOS of physical electrons is calculated by \cite{Mross2011}
\begin{eqnarray}
N(\omega)=\int d\Omega N_{I}(\Omega)N_{S}(\omega-\Omega)[\theta(\Omega)-\theta(\Omega-\omega)],\label{eq3}
\end{eqnarray}
where $N_{I}(\Omega)$, $N_{S}(\omega)$ are the DOS for Ising spin $\hat{\tau}^{x}_{i}$ and auxiliary fermion $f_{\sigma}$. [In Ref.[\onlinecite{Mross2011}], the DOS is calculated for the slave-rotor approach.] For the Ising spin is gapped in the $Z_{2}$ BCS phase, its DOS can be
approximated as $N_{I}(\Omega)\simeq \theta(\Omega-\Delta)$ with $\Delta$ being the gap of Ising spin. In the disordered states of Ising spins, its Green function reads
$G_{I}(q,\Omega)\sim \frac{1}{\Omega^{2}-q^{2}-\Delta}$.\cite{Zhong2012} One integrates it over momentum $q$ and can obtain the DOS. The DOS of auxiliary fermions has the usual BCS superconducting expression, which reads
\begin{eqnarray}
N_{s}(\omega)=N_{F}\frac{|\omega|}{\sqrt{\omega^{2}-\Delta_{SC}^{2}}}\theta(|\omega|-\Delta_{SC})\nonumber
\end{eqnarray}
for uniform s-wave pairing states and
\begin{eqnarray}
N_{s}(\omega)&&=\frac{2N_{F}}{\pi}\int dx \frac{1}{\sqrt{1-x^{2}}} \frac{1}{\sqrt{1-(\Delta_{SC}/\omega)x^{2}}}  \theta(|\omega|-\Delta_{SC})\nonumber\\
&&+\frac{2N_{F}}{\pi}\int dx \frac{1}{\sqrt{1-x^{2}}} \frac{1}{\sqrt{1-(\omega/\Delta_{SC})x^{2}}} \theta(\Delta_{SC}-|\omega|)\nonumber
\end{eqnarray}
for $d_{x^{2}-y^{2}}$-wave superconducting phase with $N_{F}$ denoting the DOS at Fermi energy. Interestingly, we obtain an analytical formalism for the electron DOS in the s-wave case as
\begin{equation}
N(\omega>0)=N_{F}\sqrt{(\omega-\Delta)^{2}-\Delta_{SC}^{2}}\theta(\omega-\Delta_{SC}-\Delta).\label{eq4}
\end{equation}

\begin{figure}
\includegraphics[width=0.9\columnwidth]{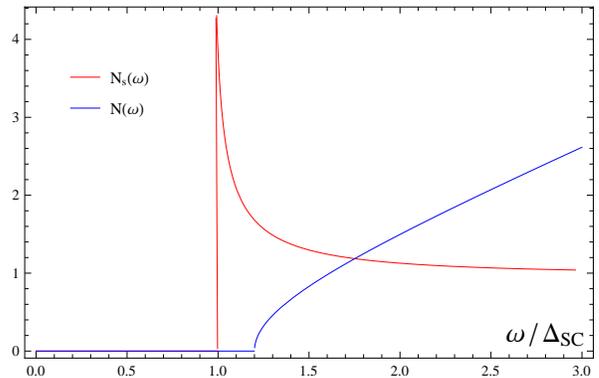}
\caption{\label{fig:1}  The local DOS of usual s-wave BCS superconducting state ($N_{s}(\omega)$) versus the $Z_{2}$ fractionalized s-wave BCS phase ($N(\omega)$) with $\Delta_{k}=\Delta_{SC}$ .}
\end{figure}

\begin{figure}
\includegraphics[width=0.9\columnwidth]{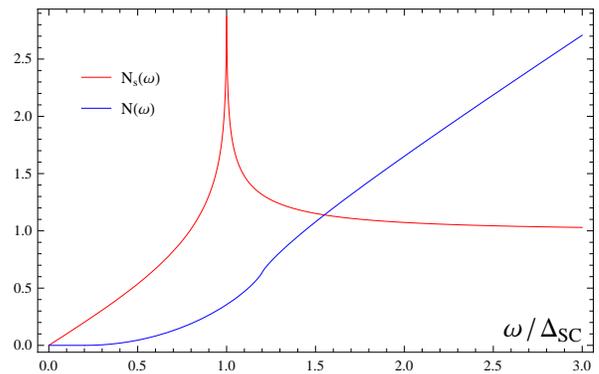}
\caption{\label{fig:2}  The local DOS of usual $d_{x^{2}-y^{2}}$-wave BCS superconducting state ($N_{s}(\omega)$) versus its $Z_{2}$ fractionalized counterpart ($N(\omega)$) with $\Delta_{k}=\Delta_{SC}(\cos(kx)-\cos(ky))$ .}
\end{figure}

We have shown the DOS of physical electrons for uniform s-wave pairing in Fig.\ref{fig:1} and $d_{x^{2}-y^{2}}$-wave in Fig.\ref{fig:2} with parameters superconducting gap $\Delta_{SC}=1$, Ising-spin gap $\Delta=0.2$
and $N_{F}=1$. From Figs.\ref{fig:1} and \ref{fig:2}, we observe that the well-known BCS coherent peak of the local DOS is absent in the $Z_{2}$ fractionalized superconducting state. This suppression of the BCS coherent peak
is due to the incoherent (gapped) Ising spin, which destroys the superconducting quasiparticle behaviors such as the mentioned BCS coherent peak.

Usually, the location of the BCS quasiparticle coherent peak implies
the value of superconducting gap particularly in realistic experiments. As for the $Z_{2}$ fractionalized BCS state, we find that the local DOS of physical electrons has a kink structure exactly at energy $\omega=\Delta+\Delta_{SC}$ as what can be seen in Figs.\ref{fig:1} and \ref{fig:2} and Eq. (\ref{eq4}). Such emergent kind structure is the reminiscent
of the whole gap formed by both Ising-spin and auxiliary-fermion. Thus, one can encode the total value of the Ising-spin gap $\Delta$ and the superconducting gap $\Delta_{SC}$ in spite of the lack of the usual BCS coherent peak. If certain realistic materials are true $Z_{2}$ fractionalized BCS state, and for the purpose of explicitly extracting the value of these two gaps, a useful method is to fit the superfluid density data with the effective one-band and/or two band BCS model\cite{Luo2005,Das2007}. After such fit, the superconducting gap $\Delta_{SC}$ can be extracted and combining with the result of local DOS, the Ising-spin gap $\Delta$ could be finally obtained.

\begin{figure}
\includegraphics[width=0.8\columnwidth]{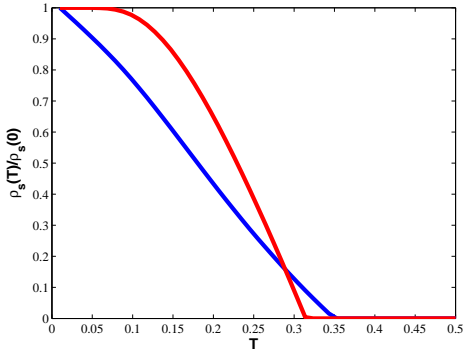}
\caption{\label{fig:3} (Color online) Normalized superfluid density $\rho_{s}(T)/\rho_{s}(0)$ of $Z_{2}$ fractionalized s-wave (red line) and $d_{x^{2}-y^{2}}$-wave (blue line) BCS superconducting state versus $T$.}
\end{figure}

For completeness, we have shown the normalized superfluid density $\rho_{s}(T)$ of the $Z_{2}$ BCS superconducting state in Fig.\ref{fig:3} (s-wave red line) and ($d_{x^{2}-y^{2}}$-wave blue line).
\begin{eqnarray}
\rho_{s}(T)=\sum_{k}\left[-\frac{\partial^{2} \varepsilon_{k}}{\partial k_{\mu}^{2}}\frac{\varepsilon_{k}}{E_{k}}\tanh(\frac{E_{k}}{2T})+2(\frac{\partial \varepsilon_{k}}{\partial k_{\mu}})^{2}\frac{\partial f_{F}(E_{k})}{\partial E_{k}}\right],\nonumber
\end{eqnarray}
when the zero-temperature superfluid density reads $\rho_{s}(0)=\sum_{k}[-\frac{\partial^{2} \varepsilon_{k}}{\partial k_{\mu}^{2}}\frac{\varepsilon_{k}}{E_{k}}]$ as normalization constant and the superconducting excitation spectrum is $E_{k}=\sqrt{\varepsilon_{k}^{2}+\Delta_{k}^{2}}$.
We should remind the reader that this normalized superfluid density totally comes from the auxiliary fermion $f_{\sigma}$ and has the identical formalism in either usual BCS or $Z_{2}$ BCS phase as expected from the study of orthogonal metal and related issues.

\subsection{The critical BCS state}
\begin{figure}
\includegraphics[width=0.8\columnwidth]{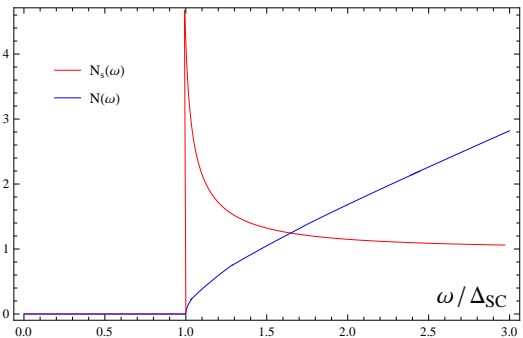}
\caption{\label{fig:5}  The local DOS of critical s-wave BCS phase ($N(\omega)$). For comparison, the result of usual s-wave BCS superconducting state ($N_{S}(\omega)$) is also shown.}
\end{figure}

\begin{figure}
\includegraphics[width=0.8\columnwidth]{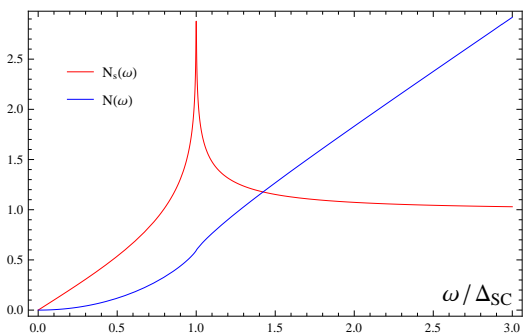}
\caption{\label{fig:6}  The local DOS of critical $d_{x^{2}-y^{2}}$-wave BCS phase ($N(\omega)$). For comparison, the result of usual $d_{x^{2}-y^{2}}$-wave BCS state ($N_{S}(\omega)$) is also shown.}
\end{figure}

Interestingly, if tuning the Ising spin part $H_{I}$ into its quantum critical point, one will have a critical BCS state following the discussion in last subsection. Note that we
do not call it critical $Z_{2}$ BCS state since the $Z_{2}$ deconfined feature of the original $Z_{2}$ gauge field is lost at criticality. Since the Ising spin is critical, its Green function may be approximated as
$G_{I}(q,\Omega)\sim \frac{1}{(\Omega^{2}-q^{2})^{1-\eta/2}}$ with $\eta\simeq0.04$ being the critical exponent of 3D Ising universal class.\cite{Zhong2012} Thus, the DOS of Ising spin reads $N_{I}(\Omega)\sim \int d^{2}q [-\frac{1}{\pi}\text{Im} G_{I}(q,\Omega)]\sim \Omega^{\eta}$. The corresponding local DOS
is shown in Figs.\ref{fig:5} and \ref{fig:6} for s-wave and $d_{x^{2}-y^{2}}$-wave pairing structure, respectively. Without surprise, we find that the BCS coherent peak is lost and there exists a kink structure at $\omega=\Delta_{SC}$ in the critical BCS state. Therefore, the qualitative physics is similar to the result of the $Z_{2}$ BCS state, however we do not expect any topological order in this critical BCS state due to the gaplessness.

\section{Extended Hubbard model}\label{sec3}
After discussing the above exactly soluble but artificial model for the $Z_{2}$ fractionalized BCS state, here we move to more realistic models like the extended $t-U-J$ model \cite{Xiang2009,Laughlin2002,Laughlin2013} to see whether the $Z_{2}$ BCS state could appear or not. The model Hamiltonian reads
\begin{eqnarray}
H=\sum_{ij\sigma}t_{ij}c_{i\sigma}^{\dag}c_{j\sigma}+\frac{U}{2}\sum_{i}(n_{i}-1)^{2}+J\sum_{<ij>}\vec{S}_{i}\cdot\vec{S}_{j},\label{eq5}
\end{eqnarray}
where $n_{i}=\sum_{\sigma}c_{i\sigma}^{\dag}c_{i\sigma}$, $U$ is the on-site Coulomb energy between electrons on the same site and the $J$ term represents the antiferromagnetic interaction between nearest neighbor sites.
Following Refs.\onlinecite{Ruegg2013,Zhong2013c}, we use the $Z_{2}$ slave-spin representation to split the physical electron $c_{\sigma}$ into an auxiliary fermion $f_{\sigma}$ and an Ising-spin $\hat{\tau}^{x}$ as
$c_{i\sigma}=f_{i\sigma}\hat{\tau}_{i}^{x}$. The auxiliary fermion $f_{\sigma}$ carries all physical quantum number of electrons while the Ising-spin $\hat{\tau}^{x}$ denoting the collective coherent motion, thus no spin-charge separation is introduced at this level in contrast to the slave-boson or slave-rotor approach.\cite{Nandkishore} Utilizing the $Z_{2}$ slave-spin representation with the constraint $\hat{\tau}_{i}^{z}+1=2(n_{i}-1)^{2}$ enforced in every site, Eq.(\ref{eq5}) is transformed into
\begin{eqnarray}
H=\sum_{ij\sigma}t_{ij}\hat{\tau}_{i}^{x}\hat{\tau}_{j}^{x}f_{i\sigma}^{\dag}f_{j\sigma}+\frac{U}{4}\sum_{i}(\hat{\tau}_{i}^{z}+1)+J\sum_{<ij>}\vec{S}_{i}\cdot\vec{S}_{j},\nonumber
\end{eqnarray}
where $\vec{S}_{i}=f_{i\alpha}^{\dag}\vec{\sigma}_{\alpha\beta}f_{i\beta}$. Here, a local $Z_{2}$ gauge structure can be seen if $f_{i\sigma}\rightarrow \epsilon_{i}f_{i\sigma}$ and $\hat{\tau}_{i}^{x}\rightarrow\epsilon_{i}\hat{\tau}_{i}^{x}$ with $\epsilon_{i}=\pm1$ while the whole Hamiltonian $H$ is invariant under this $Z_{2}$ gauge transformation.

To proceed, the mean-field treatment has to be done on this Hamiltonian and one obtains
\begin{eqnarray}
&&H_{f}=\sum_{ ij\sigma}\tilde{t}_{ij}f_{i\sigma}^{\dag}f_{j\sigma}-2J\sum_{i\sigma}m_{j} \sigma f_{i\sigma}^{\dag}f_{i\sigma}-2\lambda\sum_{i}(n_{i}-1)^{2},\label{eq6}\\
&&H_{I}=-\sum_{ ij\sigma}J_{ij}\hat{\tau}_{i}^{x}\hat{\tau}_{j}^{x}+(\frac{U}{4}+\lambda)\sum_{i}\hat{\tau}_{i}^{z},\label{eq7}
\end{eqnarray}
where the mean-field parameters are defined as $\tilde{t}_{ij}=t_{ij}\langle \hat{\tau}_{i}^{x}\hat{\tau}_{j}^{x}\rangle$, $m_{j}=m(-1)^{j}=2\langle S_{j}^{z}\rangle=\sum_{\sigma}\sigma f_{j\sigma}^{\dag}f_{j\sigma}$   and $J_{ij}=-t_{ij}\sum_{\sigma}\langle f_{i\sigma}^{\dag}f_{j\sigma}\rangle$ with $\lambda$ the Lagrangian multiplier to ensure $\langle\hat{\tau}_{i}^{z}\rangle+1=2\langle(n_{i}-1)^{2}\rangle$ at each site. Obviously, $H_{f}$ remains an
interacting Hamiltonian though we have used a mean-field approximation. Since we are interested in the possible superconducting phase and have no ambition to establish the whole phase diagram of Eq.(\ref{eq7}), we could phenomenologically add a pairing term $\sum_{k}[\Delta_{k}f_{k\uparrow}^{\dag}f_{-k\downarrow}^{\dag}+\text{H.c.}]$ into it. [Actually, the antiferromagnetic $J$ term is able to induce the superconducting pairing based on the resonance-valence-bond (RVB) picture\cite{Lee2006,Anderson2004} or antiferromagnetic spin fluctuation framework\cite{Miyake1986,Scalapino1986,Emery1986}.] Therefore, the auxiliary fermion Hamiltonian $H_{f}$ has a similar expression as the one in Eq.(\ref{eq2}) but with extra antiferromagnetic spin-density-wave mean-field term (the term with $J$ below),
\begin{eqnarray}
H_{f}&&=\sum_{k\sigma}\varepsilon_{k}f_{k\sigma}^{\dag}f_{k\sigma}-2Jm\sum_{k\sigma}\sigma f_{k+Q\sigma}^{\dag}f_{k\sigma}\nonumber\\
&&+\sum_{k}[\Delta_{k}f_{k\uparrow}^{\dag}f_{-k\downarrow}^{\dag}+\Delta_{k}^{\ast}f_{-k\downarrow}f_{k\uparrow}].\label{eq8}
\end{eqnarray}

Now, it is easy to see that when the Ising spin is condensed ($\langle\hat{\tau}^{x}\rangle\neq0$), the system is in the usual BCS phase due to nonzero overlaps between physical electrons $c_{\sigma}$ and
auxiliary fermion $f_{\sigma}$ ($c\simeq\langle\hat{\tau}^{x}\rangle f_{\sigma}$).[The antiferromagnetic spin-density-wave mean-field term does not introduce interesting physics, so we will not cover it in the following discussion.] However, if the Ising spin is disordered/gapped ($\langle\hat{\tau}^{x}\rangle=0$), we still have a superconducting state but
its single-particle behaviors are suppressed due to incoherent Ising spins as what is shown in last section. As a matter of fact, this superconducting state could be just the wanted $Z_{2}$ fractionalized BCS state
as can be seen as follows.

To demonstrate the $Z_{2}$ fractionalized BCS state, we only have to verify that it is in the deconfined phase of $Z_{2}$ gauge field since it is always superconducting.
Here, it is useful to use the action formalism of Ref.[\onlinecite{Ruegg2013}] to achieve our goal, where the dynamical $Z_{2}$ gauge field $\sigma_{ij}=e^{ia_{ij}}$ is introduced to reflect the fluctuation beyond the mean-field level,
\begin{eqnarray}
&&S=S_{\tau^{x}}+S_{f}+S_{n} \nonumber\\
&&S_{\tau^{x}}=\sum_{ij}\tilde{J}_{ij}\tau_{i}^{x}e^{ia_{ij}}\tau_{j}^{x}\nonumber\\
&&S_{f}=\sum_{ij\sigma}\tilde{t}_{ij}f_{i\sigma}^{\dag}e^{ia_{ij}}f_{j\sigma}+\sum_{ij}(\Delta_{ij}f_{i\uparrow}^{\dag}e^{ia_{ij}}f_{j\downarrow}^{\dag}+\text{H.c.}) \nonumber\\
&&S_{n}=-2i\sum_{ij}n_{ij}a_{ij},\label{eq9}
\end{eqnarray}
where $a_{ij}$ is a compact U(1) gauge field and summation over the the integer-valued auxiliary field $n_{ij}$ enforces  $\sigma_{ij}=e^{ia_{ij}}$ to be a true $Z_{2}$ field.
The trivial Berry phase term of $a_{ij}$ is neglected since it does not change the universal low-energy physics. Furthermore, the Ising-spin term can also be dropped out due to gapped Ising spins in this phase.
Thus, one may integrate over auxiliary fermions but this process turns out to be not trivial. If $f$ forms the full gapped superconducting phase such as the uniform s-wave pairing state, the integration can be done without any
singularity and we have the BF term below
\begin{eqnarray}
S=\int d^{3}x \frac{1}{\pi}\epsilon^{\mu\nu\lambda}a_{\mu}\partial_{\nu}b_{\lambda}.\label{eq10}
\end{eqnarray}
Such BF term is known to well describe the universal low-energy physics of $Z_{2}$ gauge theory at its deconfined phase.\cite{Kou2008} Therefore, gaped Ising spins and s-wave pairing BCS state of auxiliary fermions indeed corresponds to the $Z_{2}$ fractionalized BCS state. But, if the superconducting phase of auxiliary fermions is gapless, we cannot perform the integration safely and exactly.\cite{Sachdev2011} This is just the terrible problem encountered in gauge-theory description of high temperature superconductivity,\cite{Lee2006} quantum magnetism\cite{Sachdev2011} and $\nu=\frac{1}{2}$ quantum Hall effect.\cite{Halperin1993} For our interested $d_{x^{2}-y^{2}}$-wave case, its low-energy excitation is the Dirac fermion at $(\pm\frac{\pi}{2},\pm\frac{\pi}{2})$ and the corresponding low-energy description reads
\begin{eqnarray}
S=\int d^{3}x [\frac{1}{\pi}\epsilon^{\mu\nu\lambda}a_{\mu}\partial_{\nu}b_{\lambda}+\sum_{\sigma}\bar{\psi}_{\sigma}\gamma_{\mu}(\partial_{\mu}-ia_{\mu})\psi_{\sigma}].\label{eq11}
\end{eqnarray}
As argued by Senthil and Fisher,\cite{Senthil2000b} the topological order (nontrivial ground-state degeneracy on closed manifold) of deconfined $Z_{2}$ gauge theory survives when the matter field has a
linearly gapless spectrum (the Dirac fermions here are belong to this case). This means that the BF term is stable even when the gauge field $a_{\mu}$ is coupled to gapless Dirac fermion $\psi_{\sigma}$. [BF term encodes the topological order and vice-versa.] Therefore, even $d_{x^{2}-y^{2}}$-wave state of auxiliary fermions leads to the expected $Z_{2}$ structure.
So, combining the cases of s- and $d_{x^{2}-y^{2}}$-wave above, we conclude that the disordered Ising spins with s- or nodal d-wave states of auxiliary fermions represents the desirable $Z_{2}$ fractionalized BCS state.
Following this logic, there may exist $Z_{2}$ fractionalized BCS states with the chiral $p+ip$ or $d+id$ pairing\cite{Read2000} since those pairing structures are generically gapped or only has a small number of nodes.

For physical observable, we expect that the local DOS of $Z_{2}$ fractionalized BCS state found in the extended Hubbard model should be similar to the one in last section, in which the coherent BCS peak are absent.
The thermodynamic quantity, e.g., specific heat, and transport behavior such as the superfluid density are similar to the usual BCS superconductors.

\section{Discussion and remark} \label{sec4}
\subsection{Trial wave-function}
In the main text, we have study the possible $Z_{2}$ fractionalized BCS state in both exactly soluble models
and the extended Hubbard model, but it is interesting to ask what is the many-body wave-function of such fractionalized BCS state? Based on the study in Ref.[\onlinecite{Nandkishore}] and considering the superconducting background, we may propose the following typical wave-function:
\begin{eqnarray}
\Psi_{Z_{2}-BCS}(\textbf{r}_{1}\sigma_{1},\textbf{r}_{2}\sigma_{2},...\textbf{r}_{2N}\sigma_{2N})=\Psi_{PSF}({\textbf{r}_{i}})\Psi_{BCS}({\textbf{r}_{i}\sigma_{i}})\nonumber
\end{eqnarray}
where $\Psi_{PSF}({\textbf{r}_{i}})$ is the so-called paired boson superfluid wave function, which denotes the condensate of molecules formed by a pair of bosons.\cite{Nandkishore} $\Psi_{BCS}({\textbf{r}_{i}\sigma_{i}})$ is the standard BCS wave-function. With such wave-function in hand, one may use the variational Monte Carlo technique to find
the $Z_{2}$ fractionalized BCS state in various lattice models.

\subsection{Relation to $SC^{\ast}$}
We note that the authors of Ref.[\onlinecite{Chowdhury2013}] have studied the superfluid response of a $Z_{2}$ fractionalized superconducting state ($SC^{\ast}$). Such $SC^{\ast}$ state is formed by a BCS state of conduction electrons and
a $Z_{2}$ spin liquid of local electrons. This can be considered as a natural extension of the fractionalized Fermi liquid.\cite{Senthil2003,Senthil2004}
However, in our case we emphasize the pairing instability of orthogonal metals,\cite{Nandkishore} which is rather different from the $SC^{\ast}$.

\subsection{Perspective of the relation of $Z_{2}$ fractionalized BCS state and orthogonal metals}
From the viewpoint of the weak coupling, the superconducting state results from the pairing instability of certain parent normal (metallic) state and the usual BCS state indeed comes from the Fermi liquid state.
When it comes to the $Z_{2}$ fractionalized BCS state, one may expect this mechanism still works and some unknown pairing glues lead to the superconducting instability of parent orthogonal metals. If this is true, we can
write down the corresponding energy functional for the orthogonal metals like the usual Landau Fermi liquid,
\begin{eqnarray}
E=E_{0}+\sum_{k\sigma}\varepsilon_{k\sigma}\delta n_{k\sigma}+\sum_{k\sigma,k'\sigma'}f_{k\sigma,k'\sigma'}\delta n_{k\sigma}\delta n_{k'\sigma'}. \label{eq12}
\end{eqnarray}
Here, the energy of Ising spins are not included and the only difference from the Landau Fermi liquid is that the "quasiparticle" $\delta n_{k\sigma}$ represents the auxiliary fermion $f_{\sigma}$ but not realistic electrons.
The interaction effect between "quasiparticles" has been included in the function $f_{k\sigma,k'\sigma'}$. Then, one can add pairing interaction into it and some superconducting phases will be readily obtained.

However, if strong coupling is assumed (e.g. RVB mechanism \cite{Anderson2004}), particularly when system is approaching Mott transition point, the energy functional Eq.(\ref{eq12}) is useless due to the dominate localization tendency. Fortunately, following Ref.[\onlinecite{Mross2011}] and [\onlinecite{Zhong2011}] and using the hydrodynamical equation of Eq.(\ref{eq12}) and dual transformation, we can obtain actions similar to the ones encountered in gauge-theory description of high temperature superconductivity\cite{Lee2006} and Kondo-breakdown mechanism\cite{Senthil2004,Paul2007,Pepin2007} for heavy-fermion compounds. In this way, the physical electrons are fractionalized to three parts: Ising spin, fermionic spinon and bosonic chargon. Then, based on the framework of RVB,\cite{Lee2006} the pairing of spinons and condensation of chargon will lead to the expected $Z_{2}$ fractionalized BCS state.

\section{conclusion} \label{sec5}
In summary, we have proposed an unconventional $Z_{2}$ fractionalized BCS superconducting state. It has similar thermal and transport properties to the conventional BCS superconducting state, but its single-particle is strongly incoherent. Such incoherence leads to the absence of the BCS coherent peak of the local density of state in this $Z_{2}$ BCS state and this feature in principle could be probed by scanning tunneling microscopy or point-contact spectroscopy experiments. But, the kink structure found in the same DOS is the reminiscent
of the whole gap formed by both Ising-spin and auxiliary-fermion. The corresponding exactly soluble models that realize the desirable $Z_{2}$ fractionalized BCS state is presented. In addition, the extended $t-U-J$ model is analyzed and we find $Z_{2}$ BCS state may exist when the paring structure is fully gapped or has nodes. The prototypical wave-function of such $Z_{2}$ BCS state is also proposed and it may be useful as trial wave-function in numerical simulations. The relations among $Z_{2}$ BCS states, $SC^{\ast}$ and parent orthogonal metals are also analyzed. Since the parent orthogonal metals and the proposed $Z_{2}$ BCS phase have not been found experimentally, we expect they may be found in the exactly soluble models by cold-atom simulations in future. We hope the present work may be helpful for our understanding on the unconventional superconductivity and its relation to non-Fermi liquids.

\begin{acknowledgments}
The work was supported partly by NSFC, PCSIRT (Grant No. IRT1251), the Program for NCET, the Fundamental Research Funds for the Central Universities and the national program for basic research of China.
\end{acknowledgments}

\end{document}